\documentclass[fp,twocolumn,seceq]{jpsj3}
\usepackage{bm,times,amsmath,mathrsfs,graphicx}
\begin{document}
 
\title{
Phase Competition, Solitons, and Domain Walls \\
in Neutral--Ionic Transition Systems
}

\author{Masahisa Tsuchiizu$^1$\thanks{E-mail: tsuchiiz@s.phys.nagoya-u.ac.jp},
Hideo Yoshioka$^2$, and
Hitoshi Seo$^{3,4}$
}

\inst{$^1$Department of Physics, Nagoya University, Nagoya 464-8602,
Japan \\
$^2$Department of Physics, Nara Women's University, Nara 630-8506, Japan \\
$^3$Condensed Matter Theory Laboratory, RIKEN, Wako, Saitama 351-0198,
Japan \\
$^4$RIKEN Center for Emergence Matter Science (CEMS), Wako, Saitama 351-0198, Japan}

\recdate{September 9, 2016}

\abst{
Phase competition and excitations in the one-dimensional 
neutral--ionic transition
systems are theoretically studied comprehensively.
From the semiclassical treatment of the bosonized Hamiltonian, 
 we examine the competition among 
 the neutral (N), ferroelectric-ionic (I$_\mathrm{ferro}$) 
and paraelectric-ionic (I$_\mathrm{para}$) states. 
The phase transitions between them can become first-order when
 the fluctuation-induced higher-order commensurability potential is considered. 
 In particular, the description of the first-order phase boundary between
 N and I$_\mathrm{ferro}$ 
 enables us to analyze N--I$_\mathrm{ferro}$ domain walls. 
Soliton excitations in the three phases are described explicitly and 
their formation energies are evaluated across the phase boundaries.
The characters of the soliton and domain-wall excitations are 
  classified in terms of the topological charge and spin.
The relevance to the  experimental observations in 
 the molecular neutral--ionic transition systems is discussed.
We ascribe the pressure-induced crossover 
in tetrathiafulvalene-{\it p}-chloranil (TTF-CA) 
 at a high-temperature region to that 
from the N to the I$_\mathrm{para}$ state, 
 and discuss its consequence.
}

\maketitle
\section{Introduction}
\label{sec:introduction}

Neutral--ionic (NI) phase transition systems have been 
attracting interest over the past
 decades since the discovery of rich physical phenomena.
\cite{McConnell:1965uf,Torrance:1981ws,%
Okamoto:1991hj,%
Horiuchi:2006ej,Tomic:2015vr} 
These compounds are composed of two kinds of molecules stacked alternately, 
 where the valences of the donor (D) and acceptor (A) molecules are 
 nominally described as D$^0$A$^0$ and D$^+$A$^-$, 
 in the neutral (N) and ionic (I) states, respectively.
The material most intensively studied is
 tetrathiafulvalene-{\it p}-chloranil (TTF-CA), 
which resides in the  competing 
 region of the two states and exhibits the NI phase
 transition \cite{Torrance:1981ws}
 by changing temperature and/or pressures.

\begin{figure}[t]
\begin{center}
\includegraphics[width=7.5cm]{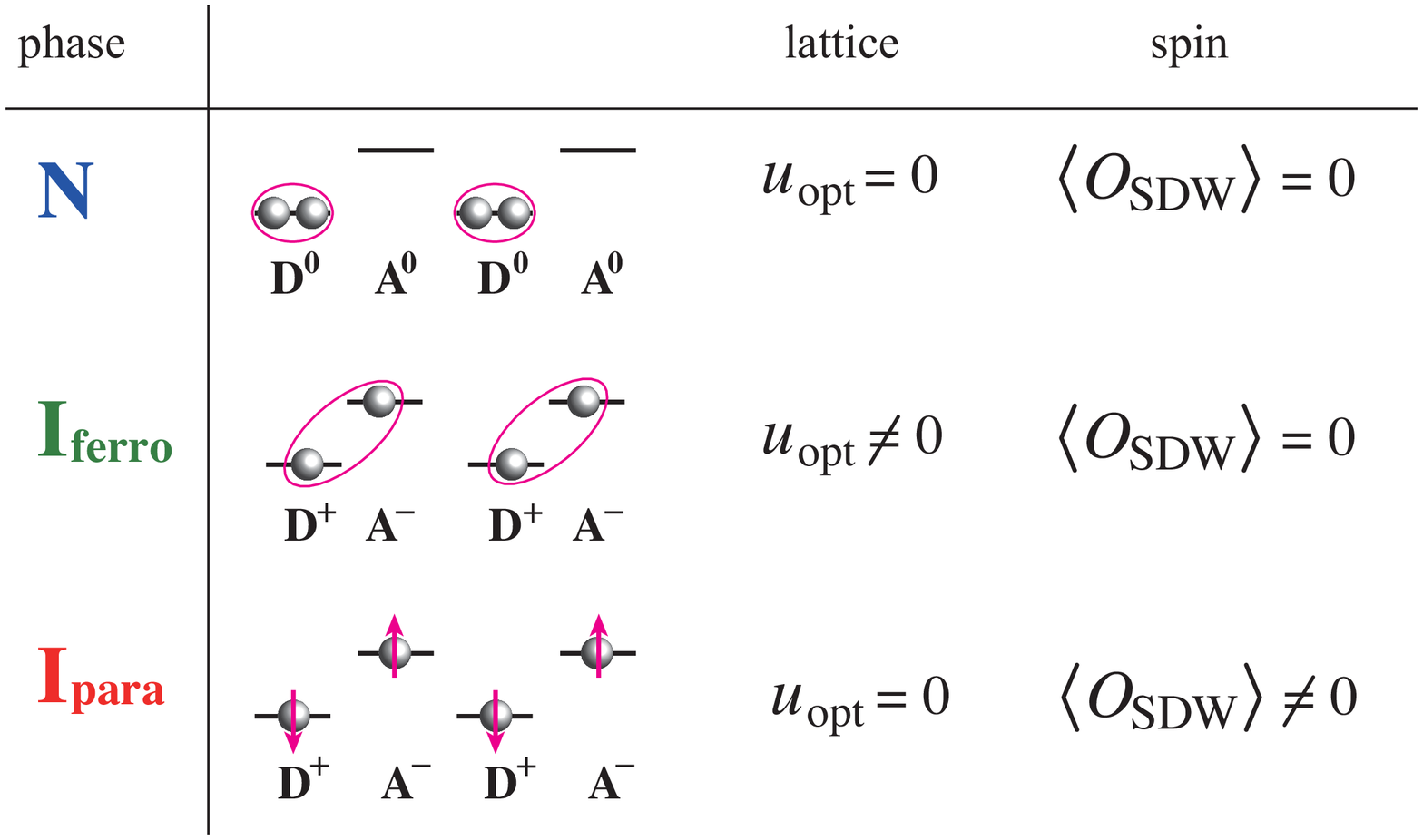}
\end{center}
\caption{
(Color online)
Schematic view of three states observed in the NI transition systems.
D and A stand for the donor and acceptor molecules, respectively,
where the HOMO of the former and the LUMO of the latter are drawn.
The red ellipses in the N and I$_\mathrm{ferro}$ states
   represent the singlet formation of the spins.
In the I$_\mathrm{para}$ state, 
the semiclassical picture of the spin degree of freedom is
  the antiferromagnetic ordering, i.e., 
 spin-density-wave (SDW). 
In the I$_\mathrm{ferro}$ state, the lattice dimerization becomes finite,
 while in the I$_\mathrm{para}$ state the SDW order parameter becomes finite
 in the semiclassical picture.
}
\label{fig:assign}
\end{figure}

The phase diagram on the plane of temperature and 
 pressure was experimentally determined, 
\cite{Mitani:1987,LemeeCailleau:1997gv,Dengl:2014aa}
in which three different states have been assigned:
the ferroelectric-ionic (I$_\mathrm{ferro}$) state, 
the paraelectric-ionic (I$_\mathrm{para}$) state, and 
the N state. 
Schematic illustrations of these states are 
depicted in Fig.\ \ref{fig:assign}.
In the N state, both the D and A sites become closed-shelled
 and thus the electronic spin degree of freedom is inactive; 
it is a band insulator (BI).
In the I$_\mathrm{ferro}$ state, 
 the lattice is dimerized and spin-singlets are formed 
   between electrons on D and A.
In this case, the dimers 
  formed by  the D$^+$ and A$^-$ molecules 
can be regarded as electric dipoles
and thus the system shows ferroelectric properties.
Then the I$_\mathrm{para}$ state, 
in which there is no static lattice dimerization, 
is in a paraelectric state. 
One can assign this state as a Mott insulator 
 with an activated spin degree of freedom. 
In TTF-CA, the I$_\mathrm{ferro}$ phase locates at low temperatures 
in the entire pressure range,
 whereas a crossover from N to I$_\mathrm{para}$ is seen 
at high temperatures with an increase in pressure.
Note that, in actual materials, 
 the valence becomes partial as D$^{+\rho}$ and A$^{-\rho}$ in all the phases, 
 owing to the quantum nature of electrons, i.e., 
the hopping integrals between sites. 
The NI transition/crossover is characterized 
 by the sudden change in $\rho$.

 One of the key factors in the NI transition systems is
 the emergence of the multi-stable structure in the free energy, 
 coupled to lattice deformation.
This is seen in the first-order nature of the 
N-to-I$_\mathrm{ferro}$ transition;
 near there, characteristic domain walls between the two states are stabilized 
 and contribute to the physical 
phenomena\cite{Mitani:1987,Okamoto:1991hj,Matsuzaki:2005fd}. 
Moreover, 
related to this fact,
it has been recognized that the various states can be controlled by 
 photo-irradiations, and that
 TTF-CA is one of the best target materials realizing 
photo-induced phase transitions.
\cite{Iwai:2002ip,Okamoto:2004wo,Tokura:2006ff,Iwai:2006jk,Uemura:2010bb}

Theoretical analyses of
the phase competition and excitations in the NI transition systems 
 have been performed in detail by Nagaosa and Takimoto
applying Monte Carlo calculations to a microscopic model 
 for the electron-lattice coupled system
\cite{Nagaosa:1986tk,Nagaosa:1986ur,Nagaosa:1986tb}. 
In particular,  the fundamental excitations are 
interpreted as soliton formations that can be described by
  the phase-field description of the order parameters.
\cite{Hara:1983vu,Fukuyama:1985jb,Nagaosa:1986tb}
Various quantum calculations  have been performed so far.
\cite{Egami:1993wt,Ishihara:1994vf,Yonemitsu:2002wn,Kishine:2004,Soos:2007,
Fabrizio:1999wl,Tsuchiizu:2004ct,Otsuka:2012}
Nevertheless, there are still issues to be clarified: 
For example, the description of the three phases seen in the experiments 
 and their competitions from a microscopic point of view 
needs to be elaborated, 
 since early works mainly focus on low-pressure states, 
namely, N and I$_\mathrm{ferro}$. 
Another important aspect is the multi-stable character stressed above, 
 which is realized numerically but has not been reproduced from 
the analytical phase-field approach.
These are the points that we discuss in this paper.

Another motivation here is the recent 
 experimental progress in high-pressure measurements on TTF-CA. 
\cite{Takehara:2015hg,Takehara:2014wg}
As a function of pressure in the high-temperature phases, 
 it has been observed that
the resistivity exhibits a characteristic suppression\cite{Mitani:1987} 
around the crossover region between the N and  I$_\mathrm{para}$ states,
 at which the spin excitations also show an abrupt change 
 in behaviors. 
A theoretical analysis 
of soliton excitations
has been performed
motivated by these findings.
\cite{Fukuyama:2016io}

In this paper, we perform a comprehensive theoretical analysis of the 
  phase competitions and excitations, 
on the basis of a microscopic interacting electron model
taking into account the electron--phonon coupling,
 by paying special attention to the multi-stable properties.
We follow the semiclassical analysis developed in Refs.\
  \citen{Hara:1983vu,Fukuyama:1985jb,Tsuchiizu:2004ct}, and
  \citen{Fukuyama:2016io}.
In Sect.\ \ref{sec:model}, we introduce the model and apply the bosonization 
 method to obtain the phase-variable description 
  of the Hamiltonian.
In Sect.\ \ref{sec:GS}, 
the ground-state properties of the semiclassical phase Hamiltonian
 are analyzed 
  by searching the potential minima.
In Sect.\ \ref{sec:soliton}, the soliton and domain excitations are analyzed.
A summary and discussions are given in Sect.\ \ref{sec:summary}.

\section{Model Hamiltonian and Bosonization}
\label{sec:model}

We consider the standard one-dimensional 
 Hubbard-type model for NI transition systems such as TTF-CA,
\cite{Nagaosa:1986tk,Nagaosa:1986ur,Nagaosa:1986tb,Rice:1982wh}
 which can describe all the three states shown in Fig.\
 \ref{fig:assign}.
The Hamiltonian is given by 
$H=H_{\mathrm{e}}+H_{\mathrm{e\!-\!ph}}+H_{\mathrm{ph}}$:
\begin{subequations}
\begin{eqnarray}
H_{\mathrm{e}} 
\!\!\! &=&  \!\!\!
-t \sum_{j,\sigma} 
\left(c_{j,\sigma}^\dagger c_{j+1,\sigma}+\mathrm{H.c.} \right)
\nonumber \\ &&{}
+ \Delta \sum_{j,\sigma} (-1)^j n_{j,\sigma}
\nonumber \\ &&{}
+ U \sum_{j} 
 n_{j,\uparrow} n_{j,,\downarrow} 
+ V \sum_{j} n_{j} n_{j+1},
\label{eq:H}
\\
H_{\mathrm{e\!-\!ph}}
\!\!\! &=&  \!\!\!
g_P
\sum_{j,\sigma}
u_j
 (-1)^j
\left(c_{j,\sigma}^\dagger c_{j+1,\sigma}+\mathrm{H.c.} \right),
\quad
\\
H_{\mathrm{ph}}
\!\!\! &=&  \!\!\!
 \frac{K_{\mathrm{ph}}}{2} \sum_{j} u_{j}^2,
\end{eqnarray}
\end{subequations}
where $c_{j,\sigma}$ is the annihilation operator of electron
on the $j$-th site with spin $\sigma$, and 
 $n_{j,\sigma}=c_{j,\sigma}^\dagger c_{j,\sigma}^{}$.
The site-alternating potential reflecting the
alternation of the D and A molecules along the chain direction
 is represented by $\Delta$.
We assume $\Delta>0$ without losing generality.
The couplings $U$ and $V$ represent the on-site and nearest-neighbor
Coulomb interactions, respectively; $H_{\mathrm{e}}$ is sometimes called 
the ionic (extended) Hubbard model.
The quantity $u_j$ represents the lattice distortion; 
 $H_{\mathrm{e \!-\! ph}}$ is the electron--phonon coupling term and 
the parameter $K_\mathrm{ph}$ is the spring constant.
Here, we consider the classical phonon and 
 focus only on the bond-alternating ($q=\pi$) distortion
 \cite{Rice:1982wh}.
The average electron density is set as half-filling, 
taking into account the HOMO 
of D and the LUMO of A molecules, 
 as shown in Fig.\ \ref{fig:assign}.

\subsection{Bosonized Hamiltonian}

Here, we represent the Hamiltonian in terms of bosonic phase-field 
variables\cite{Fukuyama:1985jb}.
By setting the lattice constant as $a$,
the charge and spin density operators can be represented as
\cite{Tsuchiizu:2002eg,Tsuchiizu:2004ct}
\begin{subequations}
\begin{eqnarray}
\frac{n_j}{a}
\!\!\! &\propto& \!\!\!
\sin(2k_Fx+\theta)\cos \phi,
\\
\frac{S_j}{a}
\!\!\! &\propto& \!\!\!
\cos(2k_Fx+\theta)\sin \phi,
\end{eqnarray}
\end{subequations}
where $\theta$ ($\phi$) represents the charge (spin) phase.
The Fermi momentum is $k_F=\pi/(2a)$.

The Hamiltonian density of the electron part is given by
\cite{Tsuchiizu:2004ct}
\begin{subequations}
\begin{eqnarray}
\mathcal H_{\mathrm{e}}
&=&
\frac{v_\rho}{4\pi}
\left[
 K_\rho(2\pi \Pi_\theta )^2
+
 \frac{1}{K_\rho}(\partial_x \theta )^2
 \right]
\nonumber \\ && {}
+
\frac{v_\sigma}{4\pi}
\left[
 K_\sigma(2\pi  \Pi_\phi )^2
+
 \frac{1}{K_\sigma}(\partial_x \phi )^2
 \right]
\nonumber \\  && {}
- \frac{g_\Delta}{2\pi^2 a^2} \sin \theta \, \cos \phi
\nonumber \\  && {}
- \frac{g_c}{2\pi^2 a^2} \cos 2\theta
 + \frac{g_s}{2\pi^2 a^2} \cos 2\phi
\nonumber \\ && {}
- \frac{g_{cs}}{2\pi^2 a^2}  \cos 2\theta \cos 2\phi
,
\label{eq:H_boson}
\end{eqnarray}
where  $K_\rho$ and $K_\sigma$ are the Tomonaga--Luttinger parameters
for the charge and spin degrees of freedom.
The operators $\Pi_\theta$ and $\Pi_\phi$ are 
  phase fields conjugate to $\theta$ and $\phi$, respectively.
The couplings are given as 
$g_\Delta=4\pi \Delta a$, $g_c=(U-2V)a+O(U^2,UV,V^2)$, 
$g_s=(U-2V)a+O(U^2,UV,V^2)$, and $g_{cs}=-2Va+O(U^2,UV,V^2)$. 
The Peierls-type electron--phonon coupling term and the lattice Hamiltonian 
 densities read
\begin{eqnarray}
\mathcal H_{\mathrm{e \! - \! ph}}
\!\!\! &=& \!\!\!
-\frac {u g_\delta}{2\pi^2 a^2} \, \cos \theta \cos \phi
,
\\
\mathcal H_{\mathrm{ph}}
\!\!\! &=& \!\!\!
\frac{K_{\mathrm{ph}}}{2a} u^2 
,
\end{eqnarray}
\end{subequations}
where $g_\delta=8\pi g_P a$.

As pointed out by Sandvik et al.\cite{Sandvik:2004js}, 
  higher-order commensurability can be important in altering the 
character of phase transitions.
 The corresponding term is expressed as
\begin{eqnarray}
\mathcal H_{\mathrm{e}}'
=
-\frac {g_{2c}}{2\pi^2 a^2} \, \cos 4\theta ,
\label{eq:He-prime}
\end{eqnarray}
where $g_{2c}>0$. 
This term effectively represents the four-body interaction,
which is absent 
in the microscopic Hubbard-type Hamiltonian; however,
it can be generated through the renormalization-group (RG) procedure.
The RG equation for $g_{2c}$ is given by
\begin{eqnarray}
\frac{d}{dl} g_{2c}
=
(2-8K_\rho) g_{2c}
+ c g_c^2 + c' g_{cs}^2 ,
\label{eq:RG}
\end{eqnarray}
where 
$l$ is the scaling parameter of the short-distance cutoff 
($a\to a\, e^{dl}$), and
$c$ and $c'$ are positive numerical constants.
Since the couplings $g_c$ and $g_{cs}$ are generally finite,
 the $g_{2c}$ term can become finite with a \textit{positive} sign 
 [$g_{2c}(l)>0$] through the RG procedure $l \rightarrow \infty$.
Eventually, this term can be relevant 
when $K_\rho<1/4$ and 
can alter the second-order phase transition into the 
first-order one
\cite{Sandvik:2004js}.
This could be confirmed by analyzing the RG explicitly; however, 
 we assume this term to be a priori
 throughout this paper,
since as we will see below it drives 
the phase transition 
between N and I$_\mathrm{ferro}$ 
to the first order, consistent with results of the experiments.
\cite{LemeeCailleau:1997gv}

Semiclassical analysis is performed
  by neglecting the phase fields 
$\Pi_\theta$ and $\Pi_\phi$.
Then the semiclassical Hamiltonian is 
\begin{eqnarray}
\mathcal{H}_{\mathrm{cl}}
\!\!\! &=& \!\!\!
\frac{v_\rho}{4\pi}
 \frac{1}{K_\rho}(\partial_x \theta )^2
+
\frac{v_\sigma}{4\pi}
 \frac{1}{K_\sigma}(\partial_x \phi )^2
\nonumber \\  && {}
+ \frac{1}{2\pi^2 a^2}
V(\theta,\phi,u).
\label{eq:classical_H}
\end{eqnarray}
The potential term is given by
\begin{eqnarray}
V(\theta,\phi,u)
\!\!\! &=& \!\!\!
- g_\Delta \, \sin \theta \, \cos \phi
\nonumber \\  && {} \!\!\!
- g_c \, \cos 2\theta
+ g_s \,  \cos 2\phi
- g_{2c} \, \cos 4\theta
\nonumber \\ && {} \!\!\!
- g_{cs}  \cos 2\theta \cos 2\phi
\nonumber \\&& {} \!\!\!
- u g_\delta \, \cos \theta \cos \phi
+ \frac{\bar K_{\mathrm{ph}}}{2} u^2 ,
\label{eq:classical_pot}
\end{eqnarray}
where we have set
\begin{eqnarray}
\bar K_{\mathrm{ph}} \equiv 
2\pi^2 a^2  \frac{K_{\mathrm{ph}}}{a}.
\end{eqnarray}
We will investigate its ground-state properties and excitations 
 in Sects.\ \ref{sec:GS} and \ref{sec:soliton}, 
 respectively.

\subsection{Order parameters}
\label{subsec:order-param}

In order to characterize the 
N, I$_\mathrm{ferro}$, and I$_\mathrm{para}$
states in the following, let us introduce order parameters.
Following the arguments in 
Ref.\ \citen{Tsuchiizu:2004ct},
we can consider the spin-density-wave (SDW),
 bond-charge-density-wave (BCDW), and 
  bond-spin-density-wave (BSDW) states.
They are characterized by the following order parameters:
\begin{subequations}
\begin{eqnarray}
\mathcal{O}_\mathrm{SDW} 
\!\!\!&\equiv&\!\!\! 
(-1)^j \,   (n_{j,\uparrow}-n_{j,\downarrow})
,
\\
\mathcal{O}_\mathrm{BCDW} 
\!\!\!&\equiv&\!\!\! 
(-1)^j 
     (c^\dagger_{j,\uparrow}c_{j+1,\uparrow}^{}
      +c^\dagger_{j,\downarrow}c_{j+1,\downarrow}^{}
      +\mathrm{H.c.})
, \quad\quad
\\
\mathcal{O}_\mathrm{BSDW} 
\!\!\!&\equiv&\!\!\! 
(-1)^j
     (c^\dagger_{j,\uparrow}c_{j+1,\uparrow}^{}
      -c^\dagger_{j,\downarrow}c_{j+1,\downarrow}^{}
      +\mathrm{H.c.})
. \quad\quad
\end{eqnarray}
\label{eq:order_parameters}%
\end{subequations}
The BCDW order parameter corresponds to the 
   Peierls dimerization operator. 
By applying the bosonization,
   the order parameters
    are rewritten as
\cite{Tsuchiizu:2004ct}
\begin{subequations}
\begin{eqnarray}
\mathcal{O}_\mathrm{SDW}(x)
\!\!\! &\propto& \!\!\!
\cos\theta(x) \, \sin \phi(x),
\label{eq:ordersdw}
\\
\mathcal{O}_\mathrm{BCDW}(x)
\!\!\! &\propto& \!\!\!
\cos\theta(x) \, \cos \phi(x),
\label{eq:orderbcdw}
\\
\mathcal{O}_\mathrm{BSDW}(x)
\!\!\! &\propto& \!\!\!
\sin\theta(x) \, \sin \phi(x).
\label{eq:orderbsdw}
\end{eqnarray}
\label{eq:orderparam}
\end{subequations}

Furthermore, to characterize low-energy excitations,
it is convenient to introduce 
   topological charges $Q$ and $S_z$
    for the charge and spin sectors,
 which we use to classify the soliton and domain-wall
  excitations.
They are given by
\cite{Nagaosa:1986tb,Fabrizio:1999wl,Tsuchiizu:2004ct}
\begin{equation}
Q = \frac{1}{\pi} \int dx \, \partial_x \theta 
, \quad
S_z = \frac{1}{2\pi} \int dx \, \partial_x \phi
.
\label{eq:topological_charge}
\end{equation}
For example, in the noninteracting case with a finite $\Delta$,
   the lowest-energy excitation is a soliton 
  represented by trajectory lines in the ($\theta$,  $\phi$) plane, 
 connecting two neighboring minima of
   the potential consisting of only one term 
$-g_\Delta\sin\theta \, \cos \phi$.
Such an excitation carries the topological charge $Q=\pm 1$ and 
    spin $S_z=\pm 1/2$, 
corresponding to a single-particle excitation
   in the N phase.

\section{Ground-State Properties}
\label{sec:GS}

First, we consider the ground-state properties 
of the semiclassical Hamiltonian Eq.\ (\ref{eq:classical_H}),
 by assuming spatially uniform solutions.
The lattice distortion can be determined by the variational approach:
\begin{eqnarray}
u_\mathrm{opt}
=
 \frac{g_\delta}{\bar K_\mathrm{ph}} \, \cos \theta \cos \phi.
\label{eq:u-opt-uni}
\end{eqnarray}
Note that Eq.\ (\ref{eq:u-opt-uni}) 
takes the same form as the BCDW order parameter in 
Eq.\ (\ref{eq:orderbcdw}). 
By inserting Eq. (\ref{eq:u-opt-uni}) into Eq. (\ref{eq:classical_pot}),  
the potential is rewritten as
\begin{eqnarray}
V(\theta,\phi,u_\mathrm{opt})
\!\! &=&\!\!
- g_\Delta \, \sin \theta \, \cos \phi
\nonumber \\  && {}
- 
\tilde g_c
\cos 2\theta
+ 
\tilde g_s
\cos 2\phi
-  g_{2c} \cos 4\theta
\nonumber \\ && {}\!\!
-  \tilde g_{cs}  \cos 2\theta \cos 2\phi,
\label{eq:VW}
\end{eqnarray}
where the coupling constants are renormalized as
\begin{subequations}
\begin{eqnarray}
\tilde g_c &\equiv&
g_c + \frac{g^2_\delta}{8\bar K_\mathrm{ph}} ,
\\
\tilde g_s &\equiv&
g_s - \frac{g^2_\delta}{8\bar K_\mathrm{ph}} ,
\label{eq:gs-tilde}
\\
\tilde g_{cs} &\equiv&
 g_{cs} + \frac{g^2_\delta}{8\bar K_\mathrm{ph}} .
\end{eqnarray}
\end{subequations}
Here, we neglect the term $-g_\delta^2 /(8\bar K_{\mathrm{ph}})$,
  which merely gives a constant contribution.

The resulting potential Eq.\ (\ref{eq:VW}) is reduced to the same form to 
 that for the purely electronic ionic extended Hubbard model 
\textit{without} lattice distortion,\cite{Tsuchiizu:2004ct,Sandvik:2004js} 
i.e., $H_{\mathrm{e}}$ [Eq.\ (\ref{eq:H})]
 whose phase-field description is 
 $\mathcal H_{\mathrm{e}}$ [Eq.\ (\ref{eq:H_boson})] added by 
 the higher-order term $\mathcal H_{\mathrm{e}}'$ 
[Eq.\ (\ref{eq:He-prime})]. 
In Ref.\ \citen{Tsuchiizu:2004ct}, 
  $\mathcal H_{\mathrm{e}}$ was analyzed, 
including the semiclassical treatment for the ground state, 
which we can directly extend.

Let us make the correspondence between the phases discussed 
 in previous works and those observed in the NI transition systems.
 They are characterized by the 
optimized lattice distortion $u_\mathrm{opt}$
and the SDW order parameter $\langle \mathcal{O}_\mathrm{SDW} \rangle$.
The correspondence is summarized in Fig.\ \ref{fig:assign}.
In the N phase, which corresponds to the BI, 
\cite{Fabrizio:1999wl,Tsuchiizu:2004ct}
 there is neither lattice dimerization nor spin ordering, i.e.,
$u_\mathrm{opt}=0$,
$\langle \mathcal{O}_\mathrm{SDW} \rangle=0$.
The I$_\mathrm{ferro}$ 
 state is characterized by the 
finite lattice dimerization $u_\mathrm{opt}\ne 0$, i.e., 
$\langle\mathcal{O}_\mathrm{BCDW}\rangle\ne 0$,  
 while 
 the spin ordering is absent owing to the spin-singlet 
   formation: $\langle \mathcal{O}_\mathrm{SDW} \rangle=0$.
In the I$_{\mathrm{para}}$ phase, 
 the spin fluctuation develops owing to 
the absence of  the lattice distortion ($u_\mathrm{opt}=0$).
In the semiclassical picture, such a Mott insulating state 
  is simply described by the SDW ordering 
  and gives $\langle \mathcal{O}_\mathrm{SDW} \rangle\ne 0$.
If we take into account the fluctuation effects for the spin mode 
  in terms of the RG method, 
the SDW ordered phase can be regarded as the 
  paramagnetic state with predominant spin correlations.
 \cite{Tsuchiizu:2002eg,Tsuchiizu:2004ct}
Namely, the positive coupling $g_s$ becomes marginally irrelevant 
and renormalized to zero $g_s\to 0$,
indicating the absence of locking potential 
for the spin phase variable $\phi$:
 the paramagnetic gapless spin-liquid state.

When there is finite electron--phonon coupling,
 the one-dimensional Mott insulating state (i.e., the I$_\mathrm{para}$ state) 
is always unstable owing to the Peierls instability
in the ground state ($T=0$).
Thus, with decreasing temperature,
the phase transition occurs from 
the paramagnetic I$_\mathrm{para}$ state into
the state with static lattice dimerization
(I$_\mathrm{ferro}$ state) where 
spin-singlets are formed.
This is seen in the form of 
 the effective coupling for the spin mode 
$\tilde g_s$ given by Eq.\ (\ref{eq:gs-tilde}).
Once the electron--phonon coupling is introduced, 
 the effective coupling $\tilde g_s$ turns negative at sufficiently   
low energy or temperature, 
and then the effective coupling $\tilde g_s$ flows to $-\infty$.
This situation corresponds to the spin-singlet formation and then 
  the SDW correlations decay exponentially.
Therefore the I$_\mathrm{para}$ state is unstable at $T=0$ 
  for finite electron--phonon coupling.
However, this is not the case if 
  we consider the states at finite temperatures,
since the temperature plays a role of cutoff of the 
RG equation, and then the effective coupling $\tilde g_s$ can be positive.
Thus the obtained I$_\mathrm{para}$ state in our semiclassical treatment
 should be regarded not as a zero-temperature phase but 
as a finite-temperature phase, 
which in fact is the case in experiments for NI transition systems.

\begin{figure}[t]
\begin{center}
\includegraphics[width=7.5cm]{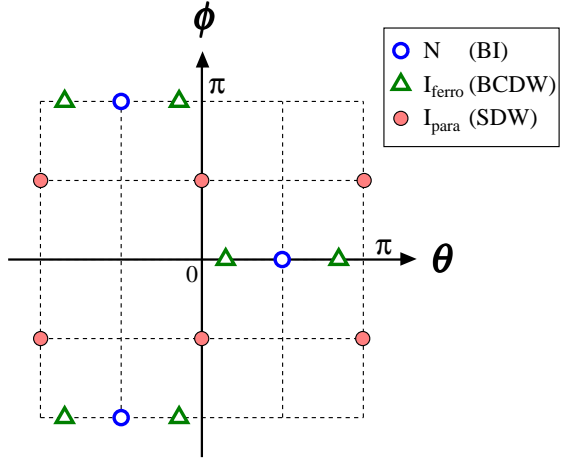}
\end{center}
\caption{
(Color online)
Positions of the locked phase fields $\theta$ and $\phi$
   in the N, I$_\mathrm{ferro}$, and I$_\mathrm{para}$ states.
The phase lockings of the N and I$_\mathrm{ferro}$ ground states 
  were originally indicated in Ref.\  \citen{Nagaosa:1986tb}.
This figure is the same as Fig.\ 8 in Ref.\ \citen{Tsuchiizu:2004ct},
where N, I$_\mathrm{ferro}$, and I$_\mathrm{para}$ correspond 
  to the BI, BCDW, and SDW states, respectively,
while the BSDW state is not shown.
In Ref.\ \citen{Fukuyama:2016io}, 
I$_\mathrm{ferro}$ is  called the polarized Mott insulator.
}
\label{fig:phase-loc}
\end{figure}

We can find the stable 
 points in the potential Eq.\ (\ref{eq:VW}), 
by assuming spatially uniform solutions.
As mentioned above, we can follow the analysis 
in Ref.\ \citen{Tsuchiizu:2004ct} 
 but here the higher-order $g_{2c}$ term $\mathcal H_{\mathrm{e}}'$ 
[Eq.\ (\ref{eq:He-prime})] is included. 
The positions of the locked phase fields $\theta$ and $\phi$ 
  are determined from the saddle-point  equations
 $\partial V(\theta,\phi,u_\mathrm{opt})/\partial \theta =
\partial V(\theta,\phi,u_\mathrm{opt})/\partial \phi = 0$.
The solutions of the saddle-point equations yield
   the following four states
($\alpha_\theta$ and $\alpha_\phi$ are nonuniversal constants), 
 which are qualitatively the same as those 
in Ref.\ \citen{Tsuchiizu:2004ct}:
\begin{itemize}
\item[(i)] N state:
The phase fields are locked at 
   $(\theta,\phi) =(\frac{\pi}{2},0)$, $(-\frac{\pi}{2},\pi)$
 (modulo $2\pi$).
In this case, $u_\mathrm{opt}=0$ and $\langle
   \mathcal{O}_\mathrm{SDW}\rangle =0$.
\item[(ii)] I$_\mathrm{ferro}$ state:
The phase fields are locked at
   $(\theta,\phi)
   =( \frac{\pi}{2}\pm \alpha_\theta,0)$ or
   $(-\frac{\pi}{2}\pm \alpha_\theta,\pi)$.
The lattice dimerization order parameter becomes finite
($u_\mathrm{opt}\propto \cos \theta \cos \phi \ne 0$) while $\langle
   \mathcal{O}_\mathrm{SDW}\rangle = 0$. 
\item[(iii)] I$_\mathrm{para}$ state:
 The phase fields  are locked at
   $(\theta,\phi)=(0,\pm \frac{\pi}{2})$ or 
   $(\pi, \pm \frac{\pi}{2})$.
The SDW order parameter becomes finite ($\langle
   \mathcal{O}_\mathrm{SDW}\rangle \propto \cos \theta \sin \phi \ne
	     0$), 
while $u_\mathrm{opt}=0$.
\item[(iv)] BSDW state:
The phase fields are locked at
   $(\theta,\phi)
    =(\frac{\pi}{2},0\pm \alpha_\phi)$ or
    $\bigl(-\frac{\pi}{2}, \pm( \pi - \alpha_\phi) \bigr)$.
\end{itemize}
The positions of the locked phase fields $\theta$ and $\phi$ and 
the corresponding states are shown in Fig.\ \ref{fig:phase-loc}.
In the present analysis, we do not focus on the BSDW phase 
  since this is not observed experimentally in TTF-CA.
Additionally, 
  the  BSDW state cannot have a true long-range order since
  the phase locking of $\phi$ is prohibited, except in the spin-gapped 
 case with $\langle \phi \rangle =0$ (mod $\pi$),
owing to the spin-rotational SU(2) symmetry.
\cite{Tsuchiizu:2004ct}

\begin{figure}[t]
\begin{center}
\includegraphics[width=7cm]{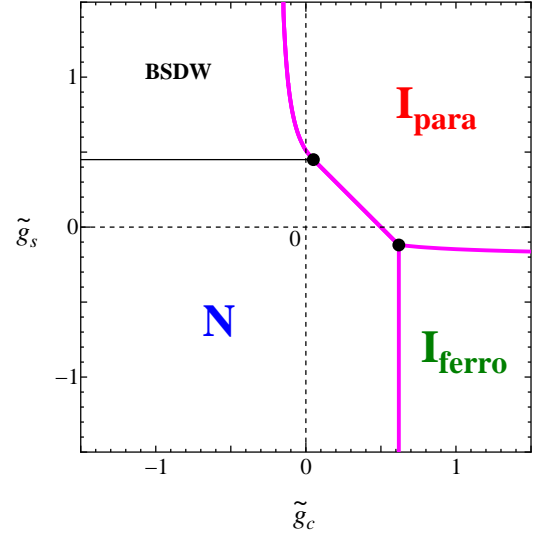}
 \end{center}
\caption{
(Color online)
Phase diagram obtained by minimizing the potential energy 
   $V(\theta,\phi,u_\mathrm{opt})$ [Eq.\ (\ref{eq:VW})].
The parameters are chosen as $g_\Delta = 1$, 
$\tilde g_{cs}=-0.2$, and $g_{2c}=0.1$.
The thick lines indicate the first-order phase transitions,
while the thin line indicates the second-order phase transition.
In the case of $g_{2c}=0$, 
 the phase boundary between N and I$_\mathrm{ferro}$ becomes 
of the second order.\cite{Tsuchiizu:2004ct}
The numerically-obtained triple point between the I$_\mathrm{para}$, 
I$_\mathrm{ferro}$, and N states
  is $(\tilde g_c, \tilde g_s) \approx (0.619,-0.119)$.
The triple point of the 
 I$_\mathrm{para}$, 
N, and BSDW states
  is $(\tilde g_c, \tilde g_s) = (0.05, \, 0.45)$.
The lattice dimerization occurs ($u_\mathrm{opt} \neq 0$) 
 in the I$_\mathrm{ferro}$ and BSDW states.
}
\label{fig:higher-comm}
\end{figure}

The analysis without the higher-order term ($g_{2c}=0$) yields the phase
diagram in Fig.\ 10 of Ref.\ \citen{Tsuchiizu:2004ct},
where the ``BI'', ``BCDW'', and ``SDW''  phases correspond to 
 the N, I$_\mathrm{ferro}$, and I$_\mathrm{para}$ states, respectively.
Owing to the presence of the $\tilde g_{cs}$ term 
that couples the charge and spin phase variables,
 the direct transition between N and I$_\mathrm{para}$ occurs 
in a finite parameter range and its phase transition 
is of the first order.
However, the transition between 
N and I$_\mathrm{ferro}$  was found to always be of the second order,
 and then the experimental observations in TTF-CA could not be reproduced.
The introduction of the $g_{2c}$ term changes it to a first-order transition. 
Once the $g_{2c}$ term is introduced, although analytical evaluations of
the phase boundaries are complicated,
here we obtain the phase diagram for a finite value of  $g_{2c}$ 
 by numerically 
finding the potential minima of $V(\theta,\phi,u_\mathrm{opt})$, 
 as shown in Fig.\ \ref{fig:higher-comm}.
We find 
that the phase boundary between the N and I$_\mathrm{ferro}$ states becomes
of the first order
in contrast to the previous analysis, 
owing to the presence of the higher-order commensurability term $\sim \cos
4\theta$.
In other words, near the first-order phase boundary, 
a multi-stable character between the
N and I states is seen, then the NI domain walls can be discussed 
as in the next section.

Incidentally, the 
 phase boundary between the I$_\mathrm{para}$ and 
I$_\mathrm{ferro}$ states also becomes of the first order.
This first-order behavior is seen even for $g_{2c}=0$ and 
can be ascribed to the spin-charge coupling 
$g_{cs}$ term. \cite{Tsuchiizu:2004ct}
As already clarified in Ref.\ \citen{Tsuchiizu:2004ct},
 this first-order behavior is an artifact due to the semiclassical treatment
 and changes into the second-order transition if we take 
 into account the fluctuation effect in terms of the RG method.
This is consistent with the spin-Peierls transition in 
 the $S=1/2$ Heisenberg chain coupled with lattice distortions,   
 where the gapless spin liquid turns into a spin-gapped dimerized state.
 It shows a second-order phase transition in general\cite{Nakano:1981vx}. 
It is also consistent with the experiments 
where the I$_\mathrm{para}$--I$_\mathrm{ferro}$
 transition as a function of temperature is continuous,
 \cite{LemeeCailleau:1997gv,Kagawa:2010hia} 
 in contrast with the first-order nature
 of the N--I$_\mathrm{ferro}$ transition.

\section{Soliton and Domain Formations}
\label{sec:soliton}

In this section, we
 examine the excited states, namely,
 the soliton and domain formations.
By taking into account the spatial variations
 of the phase fields $\theta$ and $\phi$ and 
 also of the lattice distortion $u$,  
our semiclassical Hamiltonian density [Eq. (\ref{eq:classical_H})]  is given by
\begin{eqnarray}
\mathcal{H}_{\mathrm{cl}}
\!\! &=& \!\! 
\frac{v_\rho}{4\pi}
 \frac{1}{K_\rho}[\partial_x \theta(x) ]^2
+
\frac{v_\sigma}{4\pi}
 \frac{1}{K_\sigma}[\partial_x \phi(x) ]^2
\nonumber \\  && {}
+ \frac{1}{2\pi^2a^2}
V\bigl( \theta(x),\phi(x),u(x)\bigr),
\end{eqnarray}
where the spatial variation of the lattice distortion 
is now explicitly shown [$u \to u(x)$]. 
The lattice distortion $u(x)$ can again be determined from the variational
approach:
\begin{eqnarray}
 u_{\mathrm{opt}}(x)
=
\frac{g_\delta}{\bar K_{\mathrm{ph}}}
   \cos \theta(x)\cos \phi(x).
\label{eq:u-opt-soliton}
\end{eqnarray}
In the semiclassical approach,
possible soliton/domain excitations from the ground state 
    are described as variational solutions of the following 
 simultaneous equations
\begin{subequations}
\begin{eqnarray}
0
\!\!\! &=& \!\!\!
\frac{\pi v_\rho a^2 }{K_\rho} 
\frac{\partial^2 \theta(x)}{\partial x^2}
+g_\Delta \cos \theta(x) \cos \phi(x)
\nonumber \\ && {}
-2 \tilde g_c  \sin 2\theta(x)
- 4 g_{2c} \sin 4\theta(x)
\nonumber \\ && {}
-2 \tilde g_{cs}  \sin 2\theta(x) \cos 2\phi(x),
\\
0
\!\!\! &=& \!\!\!
\frac{\pi v_\sigma a^2 }{K_\sigma} 
 \frac{\partial^2 \phi}{\partial x^2}
-g_\Delta \sin \theta(x) \sin \phi(x)
\nonumber \\ && {}
+2 \tilde g_s \,  \sin 2\phi
\nonumber \\ && {}
-2 \tilde g_{cs}  \cos 2\theta(x) \sin 2\phi(x).
\end{eqnarray}%
\label{eq:soliton-eq}%
\end{subequations}
The characters of all the possible excitations discussed 
in the following are summarized in Table \ref{table:exitations}.
In the following, we show results for the fixed condition
$g_\Delta=1$, $\tilde g_{cs}=-0.2$, $g_{2c}=0.1$,
$v_\rho=v_\sigma=2$, 
$K_\rho=0.3$, and $K_\sigma=1$.

\subsection{Soliton excitations}
\label{subsec:soliton}

\begin{figure*}[t]
\begin{center}
\includegraphics[width=15cm]{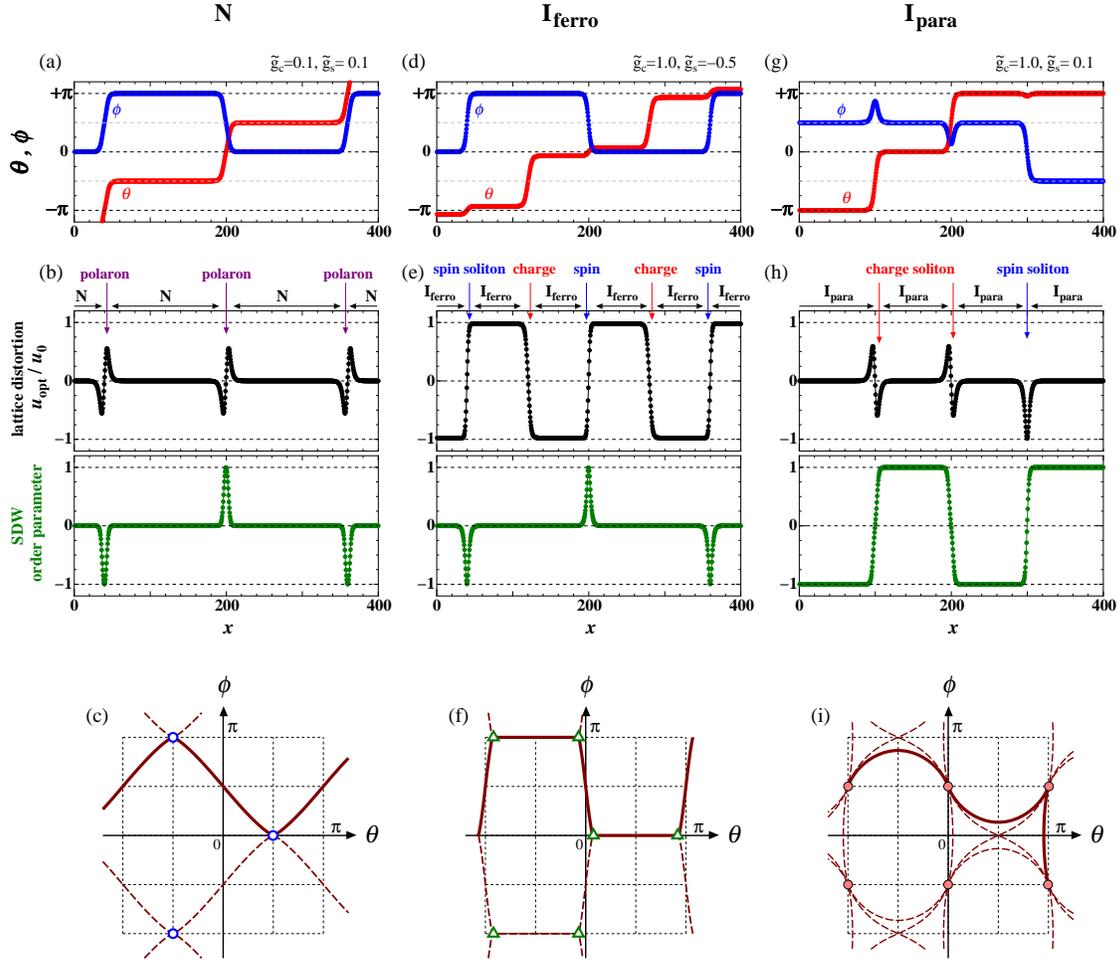}
\end{center}

\caption{
(Color online)
Soliton excitations in different phases:
 N (a-c), 
 I$_\mathrm{ferro}$ (d-f),  and 
 I$_\mathrm{para}$ (g-i). 
The parameters are chosen as 
$(\tilde g_c,\tilde g_s)=(0.1,\, 0.1)$ for the N state,
$(1.0,\, -0.5)$ for the I$_\mathrm{ferro}$ state,  and
$(1.0,\, 0.1)$ for the I$_\mathrm{para}$ state.
The spatial variations of the phase fields $\theta$ and $\phi$
are shown in (a), (d), and (g).
The variations of the order parameters $u_\mathrm{opt}$ and 
$\langle \mathcal O_\mathrm{SDW}\rangle$ are shown in (b), (e), and (h).
The corresponding trajectories on the plane of $\theta$ and $\phi$ are
shown by the solid lines in (c), (f), (i),
where the other possible trajectories are indicated by the dashed lines.
}
\label{fig:soliton}
\end{figure*}

By solving Eq.\ (\ref{eq:soliton-eq}) numerically, \cite{Tsuchiizu:2008hc}
we can analyze possible excitations.
We obtain various types of excitations depending on the 
  choice of the parameters.
The typical soliton excitations in the three phases are shown in 
Fig.\ \ref{fig:soliton}, showing the spatial variations 
of the phase fields [(a), (d), (g)] 
and the order parameters [(b), (e), (h)] as well as the trajectories in the ($\theta$, $\phi$) 
plane [(c), (f), (i)].

The soliton excitations in the N and I$_\mathrm{ferro}$ states 
are consistent with those discussed in Ref.\ \citen{Nagaosa:1986tb}.
Namely, 
in the N state, a possible soliton is the so-called polaron excitation. 
The polaron is described by the local lattice distortion and 
  carries $|Q|=1$ and $|S_z|=1/2$, adiabatically  
 connected to the single-particle excitation in the non-interacting case
 mentioned in Sect.\ \ref{subsec:order-param}.
In the I$_\mathrm{ferro}$ state, 
 two excitations are possible. 
One is the charge soliton, and the other is the spin soliton.
In both excitations,
   the topological charge $Q$ of the lowest-energy excitation
   becomes fractional,
   reflecting the nonuniversal values of the minima of 
the potential.

In the I$_\mathrm{para}$ state,
 the soliton excitation connecting potential minima along the $\theta$
 direction is the charge soliton.
Note that the charge soliton profile on the $\theta$--$\phi$
 plane largely deviates from  a straight line. 
Owing to  this fact, as seen in Fig.\ \ref{fig:soliton}(h),
it is accompanied by the
  local lattice distortion.
In this sense, this charge soliton is different from the charge
 excitation in the prototypical one-dimensional Mott insulator.
We can also consider the spin soliton, which connects the potential
 minima along the $\phi$ direction.
The spin soliton is accompanied by the 
local lattice distortion as well.
However, note that this behavior is in contrast to the charge soliton: 
In the case of the charge soliton, 
  the lattice distortion disappears if the soliton profile
  is the  straight line on the $\theta$--$\phi$ plane 
 (i.e., $\langle \phi \rangle=\pi/2$ mod $\pi$),
while the spin soliton always carries the lattice distortion even if
the spin soliton profile is described by the straight line 
(i.e., $\langle \theta \rangle=0$ mod $\pi$).
This is because of the presence of the term 
$\cos \theta(x) \cos \phi(x)$ in Eq.\ (\ref{eq:u-opt-soliton}).
We find 
$\cos \langle \phi \rangle=0$ in the case of a straight-line charge
soliton, while
$\cos \langle \theta \rangle=\pm 1$ in the case of a straight-line spin soliton.
Furthermore, 
we also observe a charge-spin coupled excitation, connecting the 
  minima, e.g., $(\theta,\phi)=(0,\pi/2)$ and $(\pi,-\pi/2)$:
This carries $|Q|=1$ and $|S_z|=1/2$. 
Its profile shown in the dashed line in Fig.\ \ref{fig:soliton}(i) 
is different from the isolated charge or spin solitons
  and then  we call this ``charge+spin'' soliton.
 When we decrease $\tilde g_c$,
the profile of 
this charge+spin soliton smoothly connects to that 
 of the charge soliton.

\begin{table}[t]
\caption{Characters of excitations:
charge soliton (CS), spin soliton (SS),
charge$+$spin soliton (CSS),
 and domain wall (DW).
The ``phase change''  indicates 
  that the pattern of the lattice distortion changes 
  across the excitation position, e.g.,
  $u_\mathrm{opt}(x)>0(<0)$ for $x>x_e$ ($x<x_e$),
 where $x_e$ is the location of the excitation.
The ``local distortion$^*$'' indicates the 
 asymmetric local lattice distortion, i.e.,
  $u_\mathrm{opt}(x_e+\delta x)= - u_\mathrm{opt}(x_e-\delta x)$.
}
\label{table:exitations}
\setlength{\tabcolsep}{1.5mm} 
{\renewcommand\arraystretch{1.2}
\begin{tabular}{lcclc}
\hline
Excitation  & Charge $|Q|$ & Spin $|S_z|$  &  Lattice distortion & Ref.
\\ \hline
Polaron in N 
& 1  & 
$1/2$
& Local distortion$^*$
& \citen{Nagaosa:1986tb}
\\
\hline
CS in I$_\mathrm{ferro}$
& $2\alpha_\theta/\pi$
& 0
& Phase change
& \citen{Nagaosa:1986tb,Fukuyama:1985jb}
\\
SS in I$_\mathrm{ferro}$
& $1-2\alpha_\theta/\pi$
& $1/2$
& Phase change
& \citen{Nagaosa:1986tb,Fukuyama:1985jb}
\\
\hline
CS in I$_\mathrm{para}$
& 1
& 0
& Local distortion$^*$
& 
\\
SS in I$_\mathrm{para}$
& 0
& $1/2$
& Local distortion
& 
\\
CSS in I$_\mathrm{para}$ 
& 1
& $1/2$ 
& Local distortion$^*$
& 
\\ 
\hline
N--I$_\mathrm{ferro}$ DW
& $\alpha_\theta/\pi$
& 0
& Zero$\leftrightarrow$Finite
\\
N--I$_\mathrm{para}$ DW
& $1/2$ 
& $1/4$
& Local distortion
\\
I$_\mathrm{ferro}$--I$_\mathrm{para}$ DW
& $1/2-\alpha_\theta/\pi$
& $1/4$
& Finite$\leftrightarrow$Zero
\\
\hline
\end{tabular}
}
\end{table}

\subsection{Soliton formation energies}

From the results above, 
we evaluate the soliton formation energies across the NI phase
boundaries, keeping in mind the experimental system 
showing the phase transition/crossover by changing temperature and pressure. 
In particular, we pay attention to the possible excitations 
contributing to the electric conductivity.

\begin{figure*}[t]
\begin{center}
\includegraphics[width=12cm]{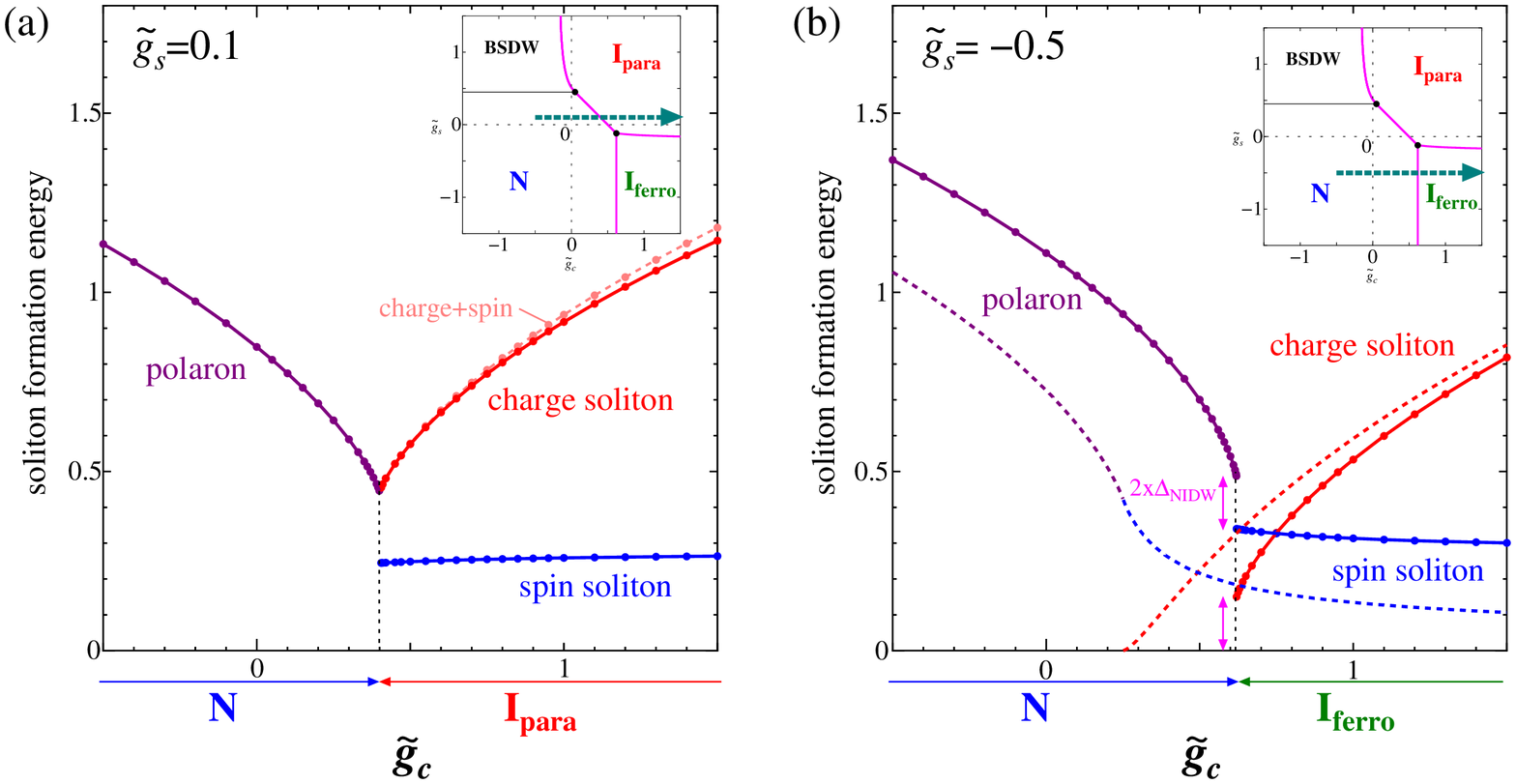}
\end{center}
\caption{
(Color online)
Soliton formation energies across the boundary between 
N and I$_\mathrm{para}$ (a) and
 the boundary between N and I$_\mathrm{ferro}$ (b).
In (b),
the case for $g_\delta=g_s=g_{cs}=g_{2c}=0$ \cite{Fukuyama:2016io}
  is shown by the dashed lines.
The difference between the excitation energies of 
a polaron and a spin soliton 
($\Delta_\mathrm{polaron}-\Delta_\mathrm{spin}$)
 at the phase boundary ($\tilde g_c\approx 0.62$),
as well as the charge-soliton excitation energy,
  corresponds to twice the excitation gap of the NIDW
(see text).
In the insets, the adopted parameters
in Fig.\ \ref{fig:higher-comm}  are shown.
}
\label{fig:soliton-formation}
\end{figure*}

The results for soliton excitations across the boundary between 
the N and I$_\mathrm{para}$ states 
  are shown in  Fig.\ \ref{fig:soliton-formation}(a).
For small $\tilde g_c$, 
  the system is in the N phase and the fundamental excitation is 
 given by the polaron.
For large $\tilde g_c$, 
  the system is in the I$_\mathrm{para}$ state and the possible
  excitations are
charge,  spin, and charge+spin solitons.
We observe that the excitation energy of 
  charge+spin soliton is always slightly larger than that of the charge soliton.
In the I$_\mathrm{para}$ state, 
the spin soliton does not carry a charge and thus 
  does not contribute to electric conductivity. 
Thus electric conduction is possible through excitations of the
 polaron soliton and the charge and charge+spin solitons 
in the N and  I$_\mathrm{para}$ phases,
 respectively.
As seen in  Fig.\ \ref{fig:soliton-formation}(a), 
all of their excitation energies become sufficiently 
suppressed, monotonically toward the 
   phase boundary. 
On the phase boundary ($\tilde g_c=0.4$), 
  the excitation energy of the polaron and 
  that of the charge soliton coincide.

The soliton excitation  energies across the boundary between 
the N and I$_\mathrm{ferro}$ states 
  are shown in  Fig.\ \ref{fig:soliton-formation}(b).
In the I$_\mathrm{ferro}$ state, both the charge and spin solitons 
are responsible for the electric conductivity,
\cite{Nagaosa:1986tb,Fukuyama:2016io}
 since both carry the topological (fractional) charge $Q$.
With increasing $\tilde g_c$, in the N phase,
the polaron excitation energy decreases.
On the other hand, in the I$_\mathrm{ferro}$ phase,
  the charge-soliton excitation energy increases,
while that of the spin soliton decreases.
Such a behavior is  seen as well in  the 
case of  $g_\delta=g_s=g_{cs}=g_{2c}=0$,
which  corresponds to the results of Ref.\ \citen{Fukuyama:2016io}. 
This is seen in the dashed lines in 
 Fig.\ \ref{fig:soliton-formation}(b):
the polaron excitation energy smoothly connects to 
  the spin-soliton excitation energy across the phase boundary.
In the present case 
 with  $g_{2c} \neq 0$,  there is a gap between them
owing to the first-order nature of the phase transition.

In both cases above, 
 just at the phase boundaries, 
 there are relations between the soliton formation energies of the two phases 
 when approaching from both sides of the critical values.
There, domain-wall excitations become possible as discussed 
in previous works, 
which we discuss in the next subsection.

\subsection{Domain walls}

In this subsection, we analyze the possible excitations 
at the phase boundaries.
At the phase boundaries, 
\textit{domain excitations} become possible 
in addition to the  soliton excitations.
In previous theoretical and experimental works,
\cite{Nagaosa:1986ur,Okamoto:1991hj,Iwai:2002ip,Okamoto:2004wo,%
Matsuzaki:2005fd,Iwai:2006jk,Horiuchi:2006ej,Soos:2007,Masino:2007df,%
Tomic:2015vr,Takehara:2015hg,Takehara:2014wg}
the domain wall between the N and I states  (abbreviated as NIDW) 
 has been discussed as a possible fundamental 
  excitation 
near the NI phase transition point.
However, the explicit descriptions of the NIDW 
  were not given in terms of the phase-field ($\theta,\phi$) description 
as mentioned in Sect.\ \ref{sec:introduction},
since 
the multi-stable character of the N and I$_\mathrm{ferro}$ states
could not be realized.
In the present analysis, in contrast, the multi-stable structure of the potential 
is realized by taking into account the higher-order commensurability 
potential 
 and thus it is now possible to give explicit descriptions of
 NIDW.

\begin{figure*}[t]
\begin{center}
\includegraphics[width=15cm]{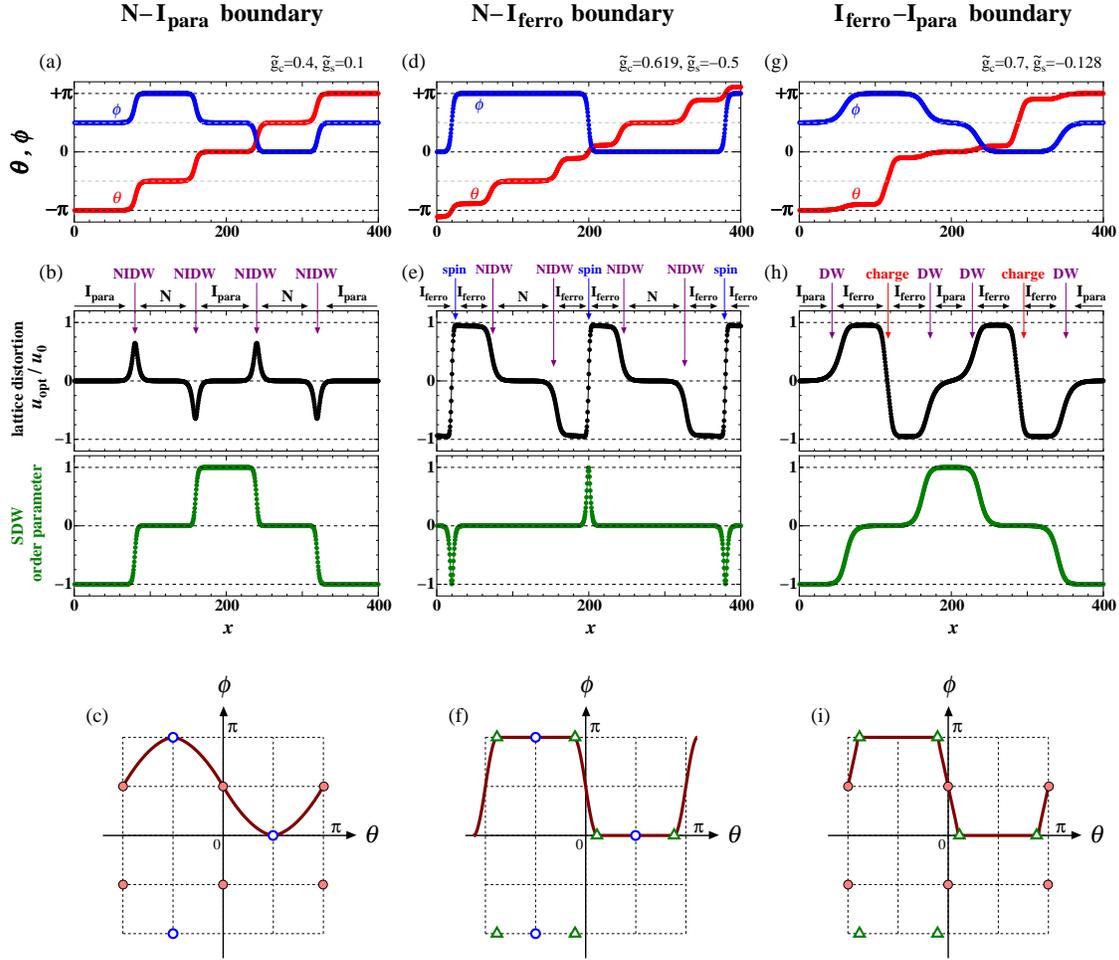}
\end{center}

\caption{
(Color online)
Domain and soliton excitations 
at the N--I$_\mathrm{para}$ boundary (a-c), 
 at the N--I$_\mathrm{ferro}$ boundary (d-f),
and at the I$_\mathrm{ferro}$--I$_\mathrm{para}$ boundary (g-i),
where the parameters are chosen as 
$(\tilde g_c,\tilde g_s)=(0.4,\, 0.1)$, $(0.619,\, -0.5)$, and 
$(0.7,\, 0.128)$, respectively.
The notations are the same as in Fig.\ \ref{fig:soliton}.
}
\label{fig:domain}
\end{figure*}

Figure \ref{fig:domain}  shows lowest-energy excitations 
on the various phase boundaries 
with the boundary conditions $Q=2$ and $S_z=0$. 
Owing to the finite value of $Q$, 
 they  can carry electronic currents when an electric field is applied.
In the following, possible excitations are analyzed 
in the respective cases.

\subsubsection{N--I$_{\mbox{\scriptsize\textit{para}}}$ boundary}

At the phase boundary between the N and I$_\mathrm{para}$ states,
the only stable excitation is the NIDW 
[Figs.\ \ref{fig:domain}(a)$-$\ref{fig:domain}(c)].
In this case, the potential $V(\theta,\phi,u_\mathrm{opt})$
takes minima at, e.g.,  
$(\theta,\phi)=(\pi/2,0)$
(corresponding to the N state) 
and $(\theta,\phi)=(0,\pi/2)$
(corresponding to the I$_\mathrm{para}$ state).
The NIDW is described by the excitation connecting these minima.
In order to distinguish this from the NIDW discussed in the next subsection, 
 we call this ``N--I$_\mathrm{para}$ DW''. 
As we will see below, this  N--I$_\mathrm{para}$ DW smoothly connects to the 
  polaron in the N phase [Fig.\ \ref{fig:soliton}(c)] or
 to the charge soliton in the I$_\mathrm{para}$ phase 
[Fig.\ \ref{fig:soliton}(i)] 
 if we move away from the phase boundary. 
This is the reason why the polaron excitation energy 
  and the charge-soliton excitation energy coincide at the 
boundary [Fig.\ \ref{fig:soliton-formation}(a)].
Here it is worth noting that
the lattice distortion occurs \textit{locally} at 
  the  N--I$_\mathrm{para}$ DW.
This is natural since the polaron in the N phase and 
 the charge soliton in the I$_\mathrm{para}$ phase 
both have such a character.

\begin{figure}[t]
\begin{center}
\includegraphics[width=8.5cm]{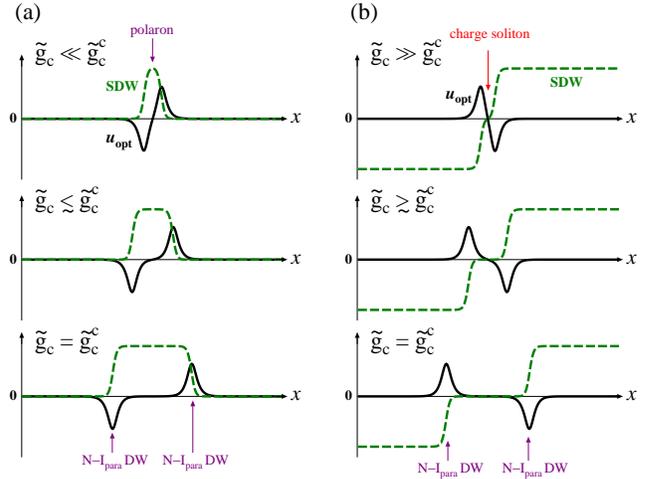}
\end{center}
\caption{
(Color online)
Generations of 
the N--I$_\mathrm{para}$ DWs
when approaching the N--I$_\mathrm{para}$ boundary
from the N state (a) and from the I$_\mathrm{para}$ state
 (b),
where we fixed $\tilde g_s=0.1$.
The solid lines denote the spatial variations of 
  $u_\mathrm{opt}$ and the dashed lines denote those of
  the SDW order parameter $\langle \mathcal O_\mathrm{SDW} \rangle$.
The polaron in the N state (a) splits into a pair of the 
N--I$_\mathrm{para}$ DWs, and 
  the region sandwiched by them is the I$_\mathrm{para}$ state 
where $\langle \mathcal O_\mathrm{SDW} \rangle \neq 0$.
The charge soliton in the I$_\mathrm{para}$ state (b) 
  splits into a pair of the N--I$_\mathrm{para}$ DWs 
and the sandwiched region is the N state
where $u_\mathrm{opt}=\langle \mathcal O_\mathrm{SDW} \rangle = 0$.
}
\label{fig:soliton-disso}
\end{figure}%

\begin{figure}[t]
\begin{center}
\includegraphics[width=8.5cm]{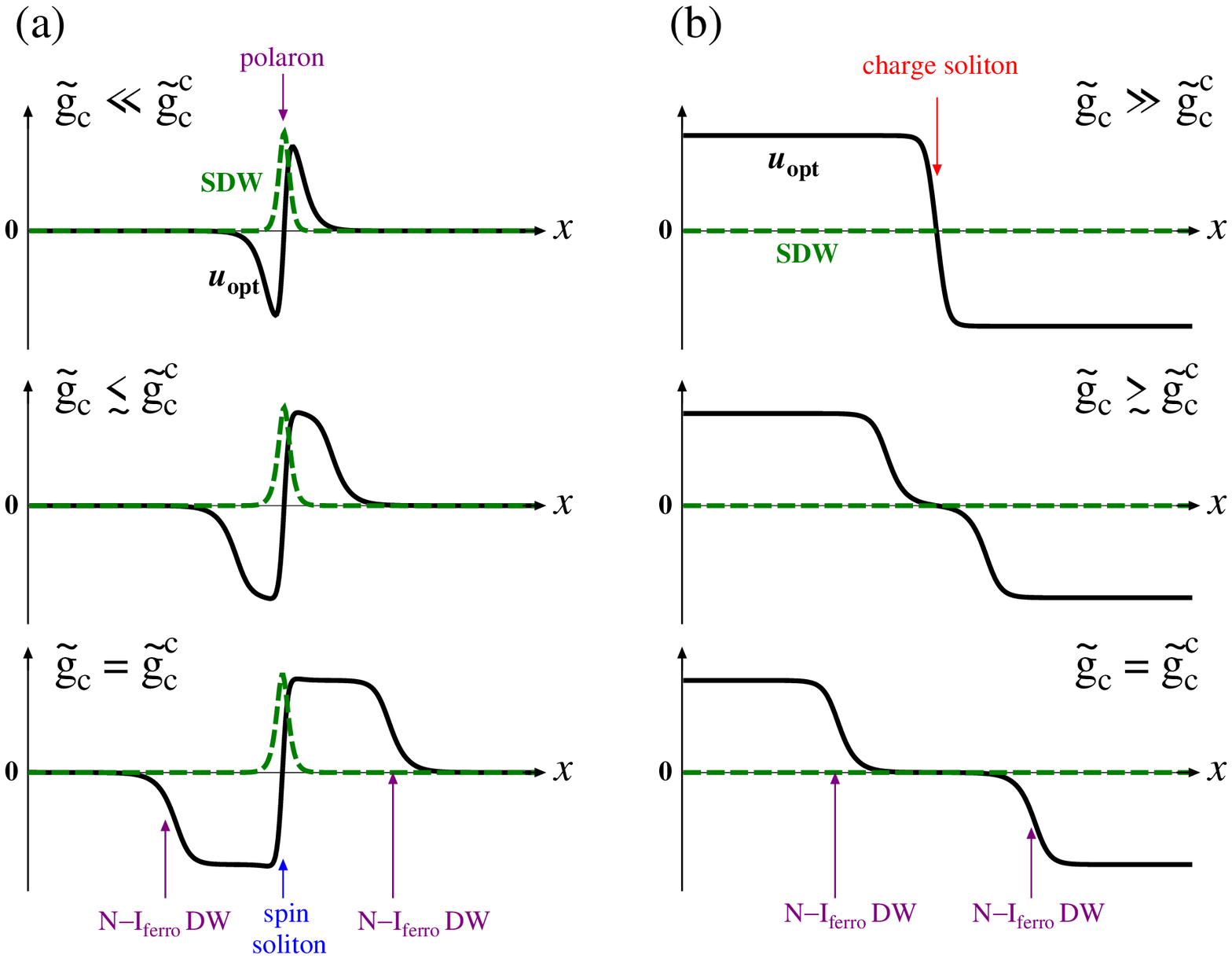}
\end{center}
\caption{
(Color online)
Generations of 
the N--I$_\mathrm{ferro}$ DWs
when approaching the N--I$_\mathrm{ferro}$ boundary
from the N state (a) and from the I$_\mathrm{ferro}$ state (b),
where we fixed $\tilde g_s=-0.5$.
The notations are the same as those in Fig.\ \ref{fig:soliton-disso}.
The polaron in the N state (a) splits into two N--I$_\mathrm{ferro}$ DWs
 and a single spin soliton, where 
the lattice-distorted region of the polaron is expanded and changes into the 
I$_\mathrm{ferro}$ state.
The charge soliton in the I$_\mathrm{ferro}$ state (b) 
  splits into a pair of the N--I$_\mathrm{ferro}$ DWs 
and the lattice-relaxed region is the N state.
}
\label{fig:soliton-disso.2}
\end{figure}%

The polaron in the N phase can be considered as a confined particle of 
 the pair of the local lattice distortion $u>0$ and $u<0$. 
The N--I$_\mathrm{para}$ DW  can be regarded as 
a dissociation of this pair of lattice distortion, i.e.,
 the single polaron is deconfined into two 
N--I$_\mathrm{para}$ DWs  at the phase boundary 
and the I$_\mathrm{para}$ state emerges as an intervening state
  between them, as shown in Fig.\ \ref{fig:soliton-disso}(a).
This picture is in accord with the observation that
the single polaron carries the charge $|Q|=1$ and 
 the spin $|S_z|=1/2$,
while 
the single N--I$_\mathrm{para}$ DW carries 
$|Q|=1/2$  and $|S_z|=1/4$.
Similarly, approaching from the  I$_\mathrm{para}$ phase, 
 the charge soliton is dissociated to a pair of N--I$_\mathrm{para}$ DW,
as shown in Fig.\ \ref{fig:soliton-disso}(b).
This picture is also in accord with 
the conservation of
the topological charge and spin.

\subsubsection{N--I$_{\mbox{\scriptsize\textit{ferro}}}$ boundary}

At the phase boundary between the N and I$_\mathrm{ferro}$ states,
 two kinds of excitations are possible 
 [Figs.\ \ref{fig:domain}(d)$-$\ref{fig:domain}(f)].
One is the NIDW, sometimes called \cite{Nagaosa:1986tb}
the lattice relaxed (LR-) NIDW,
 and the other is the spin soliton. 
In this case, the potential takes minima at, e.g.,  
$(\theta,\phi)=(\pi/2,0)$ and $(-\pi/2,\pi)$
(corresponding to the N state) 
and $(\theta,\phi)=(\pi/2 \pm \alpha_\theta,0)$ and
 $(-\pi/2\pm \alpha_\theta,\pi)$
(corresponding to the I$_\mathrm{ferro}$ state).
The NIDW is described by the excitation connecting the minima, e.g.,
$(\theta,\phi)=(\pi/2,0)$ and $(\pi/2- \alpha_\theta,0)$.
This can be called ``N--I$_\mathrm{ferro}$ DW''. 
On the other hand, the spin soliton is 
  described by the excitation connecting the minima, e.g.,
$(\theta,\phi)=(\pi/2- \alpha_\theta,0)$ and $(-\pi/2+ \alpha_\theta,\pi)$,
where we assumed $\alpha_\theta>0$.

When we approach the phase boundary from the N state,
we can observe that 
the single polaron (connecting $(-\pi/2,\pi)$ and $(\pi/2,0)$) splits
 into two  N--I$_\mathrm{ferro}$ DWs and the single spin soliton 
[see Fig.\ \ref{fig:soliton-disso.2}(a)]. 
Then at the N--I$_\mathrm{ferro}$ boundary, 
we can obtain the relation 
\begin{eqnarray}
\Delta_\mathrm{polaron}
=
2 \Delta_\mathrm{N\! - \! I_{ferro}DW}+ \Delta_\mathrm{spin},
\end{eqnarray}
where $\Delta_\mathrm{polaron}$, 
$\Delta_\mathrm{N\!-\! I_{ferro}DW}$, and $\Delta_\mathrm{spin}$
are the respective formation energies.
Therefore the gap in energy of the polaron and spin excitations 
  observed in Fig.\ \ref{fig:soliton-formation}(b)
 can be understood as $2\times \Delta_\mathrm{N\! -\! I_{ferro}DW}$.

On the other hand,
 when we approach the phase boundary from the I$_\mathrm{ferro}$
 state,
the single charge soliton 
splits into two N--I$_\mathrm{ferro}$ DWs
[see Fig.\ \ref{fig:soliton-disso.2}(b)]. 
Thus we obtain
\begin{eqnarray}
\Delta_\mathrm{charge}
=
2 \Delta_\mathrm{N\! -\! I_{ferro}DW},
\end{eqnarray}
where $\Delta_\mathrm{charge}$ is the formation energy of the 
  charge soliton.
These pictures are in accord with the neutrality of the 
topological charge $Q$ and spin $S_z$.
From Fig.\ \ref{fig:domain}(f),
we can see that the single 
N--I$_\mathrm{ferro}$ DW carries $|Q|=\alpha_\theta/\pi$ and $|S_z|=0$,
while the charge soliton in the I$_\mathrm{ferro}$ state carries 
  $|Q|=2\alpha_\theta/\pi$ and $|S_z|=0$,
and the spin soliton carries
  $|Q|=1-2\alpha_\theta/\pi$ and  $|S_z|=1/2$.

\subsubsection{I$_{\mbox{\scriptsize\textit{ferro}}}$--I$_{\mbox{\scriptsize\textit{para}}}$ boundary}

At the boundary between the I$_\mathrm{ferro}$ and I$_\mathrm{para}$ states
[Figs.\ \ref{fig:domain}(g)$-$\ref{fig:domain}(i)],
the two kinds of excitations are possible.
One is the DW between I$_\mathrm{ferro}$ and I$_\mathrm{para}$ 
(I$_\mathrm{ferro}$--I$_\mathrm{para}$ DW).
The other excitation is the charge soliton.
As seen from Fig.\ \ref{fig:domain}(i),
the I$_\mathrm{ferro}$--I$_\mathrm{para}$ DW carries the 
topological charge $Q=1/2 -\alpha_\theta/\pi$
and spin $S_z=1/4$.

We note that, as mentioned in Sect.\ \ref{sec:GS}, 
 the first-order phase boundary between I$_\mathrm{ferro}$ and I$_\mathrm{para}$
 should be replaced with a second-order one if we take into account the fluctuation effect. 
Therefore the DW here, stabilized owing to the multi-stable structure of the potential, 
 is not stabilized. 
Nevertheless, 
 if a first-order phase transition is realized beyond our model, 
 the DW structure here will be relevant.

\section{Summary and Discussion}
\label{sec:summary}

In summary, we have performed comprehensive analyses of
the competing states in the neutral--ionic transition systems,
from the semiclassical treatment of the bosonized Hamiltonian.
By taking into account the higher-order commensurability potentials,
the transition between the N state and the I$_\mathrm{ferro}$ state 
 is transformed from a second-order phase transition to a
first-order phase transition, in accordance with the experiments.
Soliton excitations have been examined explicitly and 
the soliton formation energies have been evaluated.
At the phase boundaries, we have given the explicit description of 
different types of domain-wall excitations, 
 and investigate their relations with the soliton excitations in the respective phases. 
The characters of the soliton and domain-wall excitations have been 
  classified in terms of the topological charge and spin.

Let us discuss the relevance of our results to the
experimentally-observed phase diagram
 in TTF-CA.\cite{LemeeCailleau:1997gv}
The renewed phase diagram of TTF-CA has been determined recently
on the plane of temperature and pressure, 
where the phase competitions 
among N, I$_\mathrm{ferro}$, and I$_\mathrm{para}$
were observed.\cite{Takehara:2014wg}
The competitions of these states can be reproduced in 
 Fig.\ \ref{fig:higher-comm}, 
given that the I$_\mathrm{para}$ state is described by the SDW ordered
state in our semiclassical theory.
As mentioned in Sect.\ \ref{sec:GS}, 
it is known from previous theoretical works 
that this SDW ordering becomes quasi-long ranged 
if we take the 1D quantum fluctuation effects into account,
thus the paramagnetic Mott insulating state can be  observed instead.
This paramagnetic Mott insulating state is a finite temperature phase
which is unstable toward the spin-Peierls dimerization, 
i.e., the I$_\mathrm{ferro}$ state,
as was discussed
in Sect.\ \ref{sec:GS} as well. 
The first-order N--I$_\mathrm{para}$ phase boundary is expected to
turn into a crossover at high temperatures, since there is no symmetry
breaking now 
because the corresponding Mott insulating state has no spin
ordering.
From these correspondences, 
our results are consistent with the experimental phase diagram, 
as well as with recent numerical results.
\cite{Otsuka:2012}

Next, we discuss the characteristic behavior of
  resistivity, where a sharp minimum as a function of pressure was 
 observed in the high-temperature crossover region
  between the N and I$_\mathrm{para}$ states.
 \cite{Mitani:1987,Takehara:2015hg,Takehara:2014wg}
In the low-pressure N state, 
 the lowest-energy excitation is the polaron that carries a charge. 
Therefore, the rapid decrease in soliton formation energy seen in 
Fig.\ \ref{fig:soliton-formation}
 toward the phase boundary is consistent 
 with the decrease in the resistivity in experiments.
On the other hand, as for the I$_\mathrm{ferro}$ state, 
it was claimed in Ref.\ \citen{Nagaosa:1986tb} that
the activation energy in conductivity is determined by the 
larger excitation energy of the spin or charge soliton, i.e., 
by the \textit{rate-determining step}.
In Ref.\ \citen{Fukuyama:2016io}, in contrast,
the activation energy was considered to be determined by 
the sum of excitation energies of charge and spin solitons.
If we apply these interpretations to the present results across 
the  N--I$_\mathrm{ferro}$ boundary [Fig.\ \ref{fig:soliton-formation}(b)],
both interpretations indicate  that 
the resistivity takes a minimum away from the 
N--I$_\mathrm{ferro}$ boundary. 
Instead,  the soliton formation energies across
 the  N--I$_\mathrm{para}$ boundary [Fig.\
  \ref{fig:soliton-formation}(a)] suggest
the resistivity minimum  to
coincide with the transition point between the N and I$_\mathrm{para}$ states.
This is because, in the I$_\mathrm{para}$ state, 
the low-energy excitations carrying electric current are 
the charge and charge+spin solitons, 
both showing a monotonic and steep increase 
in formation energies
by going away from the phase boundary.
The distinction of the two scenarios is possible by experimentally 
  determining the pressure where the phase transformation takes place 
 and comparing with the pressure for the resistivity minimum.
The former can be extracted, e.g., 
 by analyzing the pressure dependence of the charge transfer $\rho$, 
 although it is a crossover,
 which seems to support the latter interpretation.
\cite{Takehara:2014wg}

Here we briefly discuss the magnitude of the excitation gap
at  the boundary between the N and I$_\mathrm{para}$ states.
In TTF-CA, 
the minimum activation energy was evaluated as
 $E\approx 0.055$ eV.
\cite{Takehara:2014wg}
To make a qualitative comparison between this
activation energy 
 and the present results, we should take into account the
 quantum fluctuations in the electron parts of the Hamiltonian
and go beyond the adiabatic treatment of the lattice distortion.
In addition, the excitation gap at the boundary
sensitively depends on the magnitude of the
 dynamically-generated $g_{2c}$ term in Eq.\ (\ref{eq:RG}), 
whose qualitative estimation is difficult.
Here let us just argue about possible roles of
 the quantum fluctuations due to 
 the $\Pi_\theta$ and $\Pi_\phi$ terms in Eq.\ (\ref{eq:H_boson}), 
 and the effect of phonon dynamics.
The present semiclassical treatment of the electron part
 can be justified when $K_{\nu}\ll 1$ ($\nu=\rho,\sigma$).
It is known that 
  the soliton formation energy is reduced
when quantum fluctuations are included: 
For example, 
the soliton formation gap 
induced only by the $g_c \cos 2\theta$ term 
(simple quantum sine-Gordon model),
which is given by $\sim \sqrt{g_c}$ in the classical limit, 
 is given by $\sim g_c^{1/(2-2K_\rho)}$ 
for a general value of $K_\rho(<1)$.
\cite{gogolinbook,Zamolodchikov:1995im}
As for the spin part, the excitation gap in the I$_\mathrm{para}$ 
vanishes
if we take the 1D quantum fluctuation effects into account,
 as discussed before.
Regarding the dynamical phonon,
  we can expect that 
  the excitation gap will sufficiently be reduced if 
the gap is smaller than the Debye frequency, 
 as has been indicated in the
 context of the spin-Peierls chain.
\cite{Citro:2005bo}
All these effects can be taken into account by using the RG method,
\cite{Caron:1984ct,giambook}
 whose application to the full Hamiltonian in this case for
the quantitative evaluation of the soliton excitation gap 
 is left for future works.

We note that the 
  character of spin excitations at the N--I$_\mathrm{ferro}$ boundary
is completely different from that at the N--I$_\mathrm{para}$ boundary.
In the former, the elementary spin excitation is a spin soliton, 
 which essentially should show 
an activation behavior in the spin susceptibility in the dilute limit. 
In the latter, it will be paramagnetic excitations 
 as seen in the Heisenberg spin chain;
 therefore every D and A site carries effective 
$S=1/2$ interacting along the chain.
Nevertheless, we note that
  the temperature dependence of spin susceptibility can largely deviate
 from the Bonner--Fisher behavior, owing to the renormalization effects by 
  the terms absent in the simple Hubbard model but present here, 
 e.g., the site-alternating potential term and the electron--phonon coupling.
In fact, in tetrathiafulvalene-$p$-bromanil (TTF-BA), \cite{Kagawa:2010hia}
the spin susceptibility in the I$_\mathrm{para}$ phase shows paramagnetic behavior
  above the transition temperature to the I$_\mathrm{ferro}$ ground state, but 
does not follow the Bonner--Fisher behavior.
In addition, as shown in Sect.\ \ref{subsec:soliton}, 
 we found that the spin soliton in the I$_\mathrm{para}$ state 
 is accompanied by local lattice distortion.
In this sense, our spin soliton excitation is 
 different from the conventional spinon
  excitations, and can be regarded as  a
``spin-polaron'' excitation.  
The local/dynamic lattice distortions have actually been observed 
 in the high-temperature crossover region in TTF-CA under pressure 
in the optical measurements, \cite{Matsuzaki:2005fd}
whose relevance to such a novel excitation 
might be an interesting future problem.

Finally, let us comment on experimental observations 
of photo-induced phenomena in NI systems, 
especially in the I$_\mathrm{ferro}$ phase of TTF-CA, 
which is prominent among many.
In Ref.~\citen{Iwai:2002ip}, an I$_\mathrm{ferro}$-to-N conversion 
 into stable macroscopic N domains 
 induced by pulse optical excitations is reported. 
A notable point is that, when the temperature is just below the transition temperature $T_\mathrm{NI}$, 
 the conversion occurs irrespective of the excitation density by light, 
 whereas at low temperature, a high density is needed. 
This is consistent with the multi-stable structure of free energy discussed in this paper
 and with the stability of  N--I$_\mathrm{ferro}$ DW 
just at the transition point. 
In our calculations, when the system is deeply 
in the I$_\mathrm{ferro}$ phase,  
 DW excitations are not stable; 
they will pair annihilate into a polaron excitation 
[the inverse process of Fig.\ \ref{fig:soliton-disso.2}(a)], 
and therefore multiplication of 
 the excited N state is hampered. 
However, since the present calculations are based on classical phonons, 
 its direct application to one-dimensional ``string'' excitations implicated in experiments 
 is not straightforward, but may rather serve to describe the 
 macroscopic domain of the N state in the I$_\mathrm{ferro}$ 
background observed at 
 a later time scale of the order of $100$ ps; 
 the oscillation of such a domain is suggested in Ref.\ \citen{Iwai:2002ip}.

\section*{Acknowledgments}
  
We are grateful to 
H.\ Fukuyama, M.\ Ogata,
K.\ Kanoda, K.\ Miyagawa, and
 Y. Otsuka
for valuable discussions.
 This work was supported by Grants-in-Aid for
Scientific Research 
(Nos.\ 24740232, 25400370, 26287070, 26400377, 15K04619, and 16K05442) 
from the 
Ministry of Education, Culture, Sports, Science and Technology,
Japan.

\end{document}